\documentclass{v17neutrinos}

\def\Journal#1#2#3#4{{#1} {\bf #2}, #3 (#4)}

\def\PLB{{\em Phys. Lett.}  B}

\def\PRD{{\em Phys. Rev.} D}

\def\mco{\multicolumn}

\def\ra{\rightarrow}

\def\ko{K^0}

\def\be{\begin{equation}}
\def\ee{\end{equation}}
\def\bea{\begin{eqnarray}}
\def\eea{\end{eqnarray}}

\newcommand{\Photo}{}
\usepackage{lineno}
\begin{document}
\vspace*{4cm}
\title{NEUTRINO ASTRONOMY 2017}

\author{ T.K. Gaisser }

\address{Department of Physics and Astronomy and Bartol Research Institute\\
University of Delaware, Newark, DE 19716, USA}

\maketitle\abstracts{
This overview of neutrino astronomy emphasizes observation of
astrophysical neutrinos by IceCube and interesting limits on Galactic neutrinos
from IceCube and ANTARES.}

\vspace{-.5cm}
\section{Introduction}
Neutrinos contribute in a unique way to multi-messenger astronomy.
Being electrically neutral and weekly interacting, they reach Earth
from the entire cosmos.  They are associated with hadronic interactions
in their sources, for example $p\rightarrow \pi^+\rightarrow \mu^+\,+\,\nu_\mu$,
so they are likely to be tracers of cosmic-ray sources.  Unlike cosmic rays, 
high-energy neutrinos travel 
without deviation by magnetic fields and therefor point back to their sources.  
Unlike high-energy gamma-rays, they 
do not cascade in the Cosmic Microwave Background (CMB) or in other Extra-galactic
Background Light (EBL).  The main problem is accumulating enough events
to identify the neutrino sources.  Even with the cubic kilometer size of IceCube,
the accumulation of neutrinos likely to be of astrophysical origin is at the level of a
few tens per year.

There are three high-energy neutrino detectors currently in operation,
Baikal, ANTARES and IceCube, and there are active plans for expansion associated
with all three.  Recent results were presented at this conference
by Antoine Kouchner~\cite{Kouchner:2017rdv} for ANTARES and by Carlos 
Arg\"{u}elles~\cite{Arguelles:2017rdv}
for IceCube.  This paper will review the status of the search for both galactic
and extra-galactic neutrinos.  The basic situation is that IceCube has discovered
astrophysical neutrinos at high energy above the steeply falling background
of atmospheric neutrinos.  The discovery came first in the High-Energy Starting
Event (HESE) analysis~\cite{Aartsen:2014gkd} in which events were required to start within an inner fiducial
volume of IceCube.  The excess is also seen at a consistent level in the
upward moving sample of neutrino-induced muons~\cite{Aartsen:2016xlq}.  The all-sky search for steady
point sources in IceCube~\cite{Aartsen:2016oji}, having identified no significant excess,
sets the strongest limits on the intensity of neutrinos from a single point in the sky.  
The limits are improved at low energies in the Southern sky in the combined search
by ANTARES and IceCube~\cite{Adrian-Martinez:2015ver}. This is because
ANTARES views the Southern sky through the Earth and therefore has a much lower
background than IceCube, which sees the Southern sky through a large foreground
of atmospheric muons. 

\section{Potential sources of high-energy neutrinos}\label{subsec:potential}

It is natural to divide neutrino sources into two classes, Galactic and
extra-galactic.  Within each group, both diffuse neutrinos and neutrinos from
point sources or compact regions of the sky are expected.  

\subsection{Neutrinos of Galactic origin}
The one truly guaranteed
population of high-energy extraterrestrial neutrinos arises from interactions
of cosmic rays with gas in the disk of the Milky Way~\cite{Stecker:1978ah}.
The spectrum would be a power-law with the same spectral index as the parent
cosmic-rays until the spectrum becomes
steeper (around $100$~TeV reflecting the steepening of the parent nucleon
spectrum at the knee $\sim1$~PeV).  The expected rate is obtained by
calculating the neutrino emissivity per hydrogen atom given an assumption
about the cosmic-ray spectrum in the interstellar medium (ISM) and then
summing over the distribution of gas in the ISM.  In the simplest case
the cosmic-ray spectrum is assumed to be the same as observed at Earth
everywhere in the disk (differential index $\sim 2.7$)
and the gas density is assumed to be independent
of Galactic radius.  

Because astrophysical neutrinos are produced mainly by decays of charged pions,
the Galactic neutrino flux is closely related to the $\pi^0$ contribution
to the corresponding component of the diffuse flux of gamma-rays.
In a series of papers~\cite{Gaggero:2014xla,Gaggero:2015xza,Gaggero:2015jma} 
Gaggero~\textit{et al.} use the Fermi measurements
of gamma-rays from different regions of the Galaxy~\cite{Ackermann:2012pya}
to model cosmic-ray propagation in the inner Galaxy.  They find that
the parameter $\delta$ that characterizes the energy dependence of the
diffusion coefficient ($D\propto E^\delta$) decreases toward the center of the Galaxy
compared to its value near Earth.  This in turn implies that the parent
spectrum of the pions is harder in the inner regions of the Galaxy so that 
the predicted neutrino spectrum is increasingly higher than in the conventional model
as energy increases.

Current upper limits of both
ANTARES~\cite{Adrian-Martinez:2016fei,Albert:2017oba} 
and IceCube~\cite{Aartsen:2017ujz} are
close to the prediction of the KRA$_\gamma$ model of~\cite{Gaggero:2015xza}.  The limits from
ANTARES are comparable to those of IceCube even though it is significantly smaller
because the central region of the Galaxy is in its field of view.  
The possible contribution of a Galactic component to the HESE sample
is a topic of interest~\cite{Neronov:2016bnp}.  The IceCube analysis~\cite{Aartsen:2017ujz} 
uses its gamma-ray
templates to integrate the Galactic limits over the full sky in order
to compare directly with fits to the diffuse astrophysical flux observed by IceCube.
The conclusion is that Galactic contribution is $< 14\%$ of the astrophysical flux 
reported in~\cite{Aartsen:2015knd}.  In other words, the Galactic plane is not yet
resolved in neutrinos.  The ANTARES analysis~\cite{Adrian-Martinez:2016fei} 
reaches a similar conclusion
by comparing to HESE events~\cite{Aartsen:2014gkd} consistent with being from the Galactic
plane.

Neutrinos from sources in the galactic plane are also expected, in particular from
supernova remnants accelerating particles into nearby molecular clouds~\cite{Ahlers:2015moa}.
The IceCube analysis~\cite{Aartsen:2017ujz} includes a stacking search based on
catalogs of potential Galactic sources.

\subsection{Neutrinos of extra-galactic origin}

\begin{figure}
\includegraphics[width=0.7\linewidth]{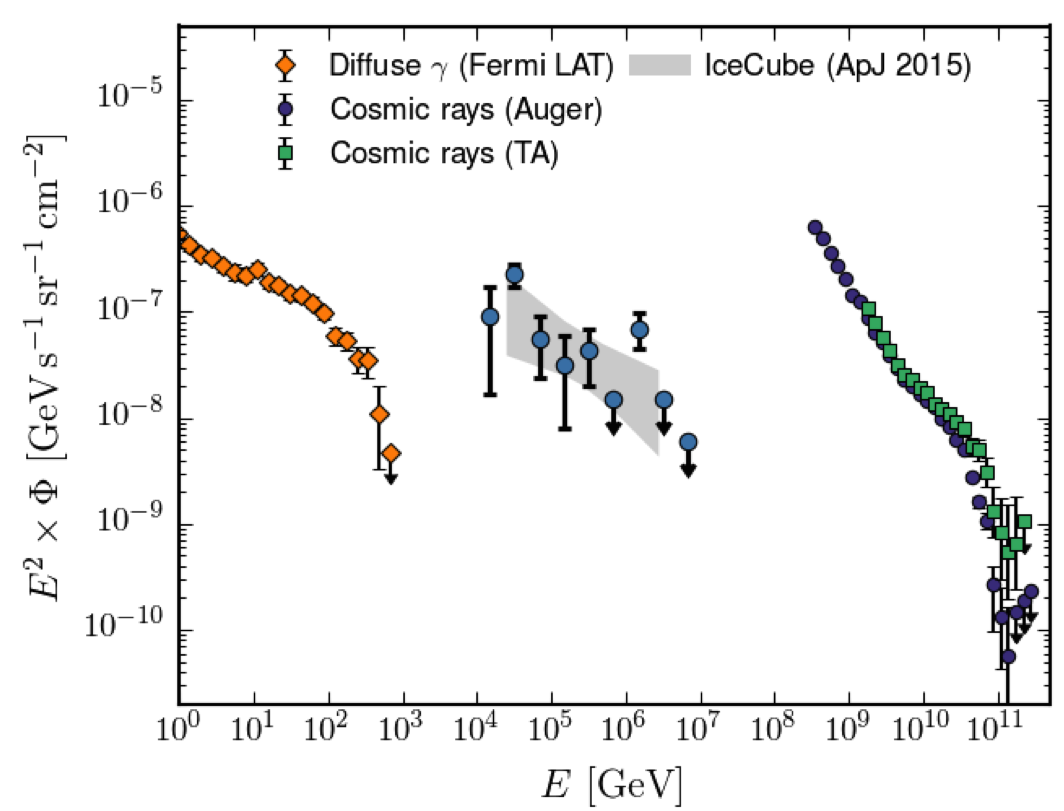}
\caption{Spectral energy distribution of the IGRB~\protect\cite{Ackermann:2014usa}, 
the IceCube neutrino flux~\protect\cite{Aartsen:2015knd},
and the UHECR~\protect\cite{ThePierreAuger:2013eja,AbuZayyad:2012ru}.
Figure adapted
from~\protect\cite{Mohrmann:2015zxq}.
}
\label{fig:multi_mess}
\end{figure}

``Cosmogenic neutrinos'' constitute a truly diffuse population of extragalactic neutrinos.
This term refers to neutrinos from decay of charged pions produced by interactions of
ultra-high energy cosmic rays (UHECR) with the CMB and EBL.  Cosmogenic neutrinos are
often referred to as ``guaranteed" because both UHECR and CMB exist.  However,
if, as suggested by Auger data~\cite{Aab:2014aea}, 
the highest energy cosmic rays are mostly nuclei rather than protons, the neutrino
flux may be so low that it is undetectable because the energy per
nucleon would be below the threshold for photo-pion production on the CMB.
IceCube limits on neutrinos
with energy greater than 10~PeV~\cite{Aartsen:2016ngq} already constrain
predictions of cosmogenic neutrinos based on the assumption that the 
highest energy cosmic-rays are protons.  The most recent limit
from nine years of IceCube data~\cite{Shigeru:2017} gives a 90\% confidence level
differential upper limit on $E_\nu^2{\rm d}N_\nu/{\rm d}E_\nu$ that is below
$2\times 10^{-8}$~GeV~cm$^{-2}$sr$^{-1}$s$^{-1}$ up to $10^9$~GeV.
Because neutrinos produced by interactions of cosmic-rays 
with energy per nucleon~$ > 10^9$~GeV would produce neutrinos with $E_\nu > 10^7$~GeV, 
this result will constrain any model that connects the IceCube
HESE flux shown in Fig.~\ref{fig:multi_mess} to the flux of UHECR.

\subsubsection{Neutrinos from cosmic-ray interactions in external galaxies}

The same processes that accelerate cosmic rays in the Milky Way are at work in 
external galaxies.  The conditions for interaction with gas to produce
neutrinos may, however, be expected to vary from one galaxy to another. 
As a starting point, it is interesting to ask what signal would be expected
from galaxies similar to the Milky Way.  The isotropic flux from a distribution
of sources with similar intrinsic luminosities $Q_\nu(E_\nu)$ ($\nu$ per sec per GeV) is
given by an integral over redshift $z$ as~\cite{Ahlers:2017lkd}
\begin{equation}\label{eq:cosmo-integral}
\phi_\nu(E_\nu) = \frac{R_H}{4\pi}\int Q_\nu((1+z)E_\nu)\,\rho(z)\,\frac{H_0}{H(z)}\,{\rm d}z
\sim \xi_z\frac{\rho_0 Q_\nu(E_\nu) R_H}{4 \pi}.
\end{equation}
Here the Hubble radius $R_H=\frac{c}{H_0}\approx 4000$~Mpc and
$\frac{H(z)}{H_0} = \sqrt{\Omega_\Lambda + \Omega_m(1+z)^3}$ with
$\Omega_\Lambda = 1-\Omega_m \approx 0.7$.  The cosmological factor
$\xi_z$, of order $1$-$3$, depends on how the sources evolve in redshift.
In his
review of galactic neutrino sources~\cite{Ahlers:2017lkd},
Markus Ahlers compares the flux calculated in Eq.~\ref{eq:cosmo-integral}
with the local flux from the Galactic plane but averaged over the full sky.  
By a dimensional argument, the latter is
\begin{equation}\label{eq:local}
\phi_{loc}\sim \frac{Q_\nu(E_\nu)}{4\pi R_{MW}^2},
\end{equation}
where the radius of the Milky Way is $R_{MW}\approx 15$~kpc.  The result is
that the extra-galactic flux is a factor of two to three orders of magnitude
less than the local flux for a density of $10^{-3}$ to $10^{-2}$ galaxies per Mpc$^3$
similar to the Milky Way.
 
In galaxies with more active star formation, the neutrino flux
should be higher because of the larger rate of supernova explosions and the 
correspondingly higher flux of cosmic rays.  In addition, 
the increased rate of turbulence may lead to a situation in which the
cosmic rays typically interact with the gas before they escape from the galaxy,
which is referred to as the ``calorimetric limit.''  The Fermi paper~\cite{Ackermann:2012vca}
shows a positive correlation between the rate of star formation and the gamma-ray
luminosity of external galaxies.  In addition, the gamma-ray spectrum tends to harden
with the rate of star formation as the spectrum of cosmic rays becomes closer to
the source spectrum.  A neutrino flux from the cumulative contribution of starburst
galaxies large enough to be observed in a kilometer scale neutrino detector was predicted
by Loeb and Waxman a decade ago~\cite{Loeb:2006tw} and discussed recently in the
review paper of Waxman~\cite{Waxman:2017lkd}.

One question that arises is whether the same sources that produce the neutrinos 
observed by IceCube also explain the origin of UHECR.  Figure~\ref{fig:multi_mess}
compares the spectral energy distribution (SED: $E^2{\rm d}N/{\rm d}E$) of three
populations of particles, the diffuse Isotropic Gamma-Ray Background 
(IGRB)~\cite{Ackermann:2014usa}, the IceCube astrophysical neutrino flux~\cite{Aartsen:2015knd}
and the high-energy end of the cosmic-ray spectrum~\cite{ThePierreAuger:2013eja,AbuZayyad:2012ru}.
Waxman points out that the Waxman-Bahcall neutrino bound~\cite{Waxman:1998yy,Bahcall:1999yr}
normalized to the UHECR above $10^{19.2}$~eV
and assuming an $E^{-2}$ spectrum for the extra-galactic component comes close to the
level of the of the IceCube measurement, shown as the sum of three flavors of neutrinos
and anti-neutrinos.  The UHECR themselves are not, however, the particles
that produce the neutrinos.  In the starburst model the neutrinos are produced by
 cosmic-rays accelerated within each galaxy, perhaps in explosions of supernovae,
 when they interact with gas inside the galaxy.  In this case (as in any scenario in
 which the neutrinos are produced by interactions of a spectrum of nucleons with
 ambient gas) the neutrino spectrum will extend to low energy.  Then, unless
 the spectrum is quite hard, the model implies an overproduction of diffuse gamma rays
 compared to the Fermi measurement in Fig.~\ref{fig:multi_mess}~\cite{Murase:2013rfa,Bechtol:2015uqb}.
 The problem is apparent from the figure (even before a calculation of the electromagnetic
 cascade in the CMB) given that the $\pi^0$~decay photon flux is
 similar at production to the neutrino flux.  If the neutrino flux is as steep as
 shown in the figure, then it is not possible to account for the entire neutrino
 flux in the starburst model~\cite{Senno:2015tra}.
 
 Another contribution to the neutrino flux can be expected from Type IIn supernovae
 in external galaxies~\cite{Murase:2010cu,Zirakashvili:2015mua}.  These are supernove that explode
 into the dense wind of a progenitor star.  The neutrinos are produced for a short 
 time until the supernova shock breaks through the dense shell (up to a 30 year time scale).
 The paper~\cite{Zirakashvili:2015mua} notes one event in the HESE sample as being $1.35^\circ$ from SN2005bx at z=0.03.
 In his presentation at ISVHECRI 2016, Ptuskin noted that track \#11 in the upward
 neutrino-induced muon sample~\cite{Aartsen:2016xlq} is $0.3^\circ$ from SN2005jq
 at z=0.23. However, no systematic analysis of chance probability has yet been done.  Moreover,
 a detailed calculation finds that only a fraction of the IceCube neutrino flux can be
 explained in this way~\cite{Petropoulou:2017ymv}.  

\subsubsection{Neutrinos from active galaxies}

Active galaxies with relativistic jets driven by accretion onto the central
supermassive black hole are potential sources of high-energy particles
including neutrinos.  In this case it is generally
assumed that the neutrinos arise from photo-pion production by accelerated
protons interacting with intense radiation fields in the inner regions
of the system~\cite{Murase:2014foa}.  Because of the high energy threshold
for the production process, the neutrino spectrum and the corresponding spectrum of 
photons, do not extend down to low energy, as in the case of a spectrum
of nucleons interacting with gas.  In addition, much of the diffuse gamma-background
is from unresolved blazars~\cite{TheFermi-LAT:2015ykq}.  There are
blazars consistent with being from the direction of neutrinos in the HESE
data sample~\cite{Padovani:2014bha}, as well as at least
 one case of a coincidence with a flaring episode~\cite{Kadler:2016ygj}.  However,
a systematic search of the Fermi 2LAC catalog~\cite{2011ApJ...743..171A,2015ApJ...806..144A} 
puts an upper limit
of $\sim 20\%$ on the fraction of the HESE flux arising from sources
in this catalog~\cite{Aartsen:2016lir}.  The current status of neutrinos
from active galaxies is reviewed by Murase~\cite{Murase:2017lkd}.
One possibility is that blazars with hard spectra contribute mainly at
high energy and give only a fraction of the astrophysical flux at lower
energy.  (See also~\cite{Padovani:2015mba}).

\subsubsection{Gamma-ray bursts}
Gamma-ray bursts (GRBs) are prominent candidates for the origin
of UHECR~\cite{Waxman:1995vg,Vietri:1995hs} as well as for high-energy neutrinos~\cite{Waxman:1997ti}.
Of the two main classes of gamma-ray bursts (GRBs), long bursts ($\ge 2$~s) associated
with core collapse of massive stars are the most likely
candidates for production of neutrinos~\cite{Meszaros:2017lkd}.
Acceleration may occur in shocks inside the relativistic jets driven by 
accretion onto a central remnant black hole.  As with AGN jets, the production
process is photo-pion production off intense radiation fields, so
the neutrino spectrum is characterized by an energy distribution that peaks
in the PeV range~\cite{Becker:2007sv}.

From the observational point of view, GRBs are ideal candidate sources,
providing a time stamp as well as a direction from satellite observations.
The latest IceCube study looks at more than 1000 identified GRBs
and finds no significant correlations~\cite{Aartsen:2017wea}.

\section{Interpretation of present results}

IceCube has identified a population of high-energy astrophysical neutrinos.
  Based on the
distribution of events in the sky, no more than $\sim$10\% of the observed
signal is from neutrinos from the Milky Way.  No sources have yet been
identified, either in the all-sky survey or in the catalog of potential
sources searched~\cite{Aartsen:2016oji}.  Because neutrinos pass undeviated
through the whole cosmos, it is possible to have a situation where the observed
sample of events is made up of one or two neutrinos from each of a large 
number of weak sources~\cite{Lipari:2008zf}.  The situation may be quantified
by comparing the flux from the whole sky (Eq.~\ref{eq:cosmo-integral}) with 
limits on the flux from a nearby source for each class of 
sources~\cite{Lipari:2008zf,Ahlers:2014ioa,Murase:2016gly}.  The flux
from a nearby source at distance $d\propto \rho_0^{-1/3}$ is
\begin{equation} \label{eq:nearby}
\phi^*_\nu\approx \frac{Q_\nu}{4\pi d^2}\approx Q_\nu\rho_0 d.
\end{equation}
Typical upper limits on extragalactic sources in the seven-year point source
catalog~~\cite{Aartsen:2016oji} correspond to an energy flux $< 2\times 10^{-9}$~GeV~cm$^{-2}$s$^{-1}$.
The product $Q_\nu\rho_0$ in Eq.~\ref{eq:nearby} is given by the observed diffuse flux from 
the last term of Eq.~\ref{eq:cosmo-integral}, so the point source
upper limit~[\ref{eq:nearby}] implies an upper limit on the distance
between sources and hence a lower limit on the density of sources.
For an observed three-flavor neutrino energy flux of $\sim 3\times 10^{-8}$
in the units of Fig.~\ref{fig:multi_mess}, $Q_\nu\rho_0\sim 10^{43}$ erg/Mpc$^3$
and the minimum source density is $\sim 10^{-7}$~Mpc$^{-3}$.

\begin{figure}
\includegraphics[width=0.6\linewidth]{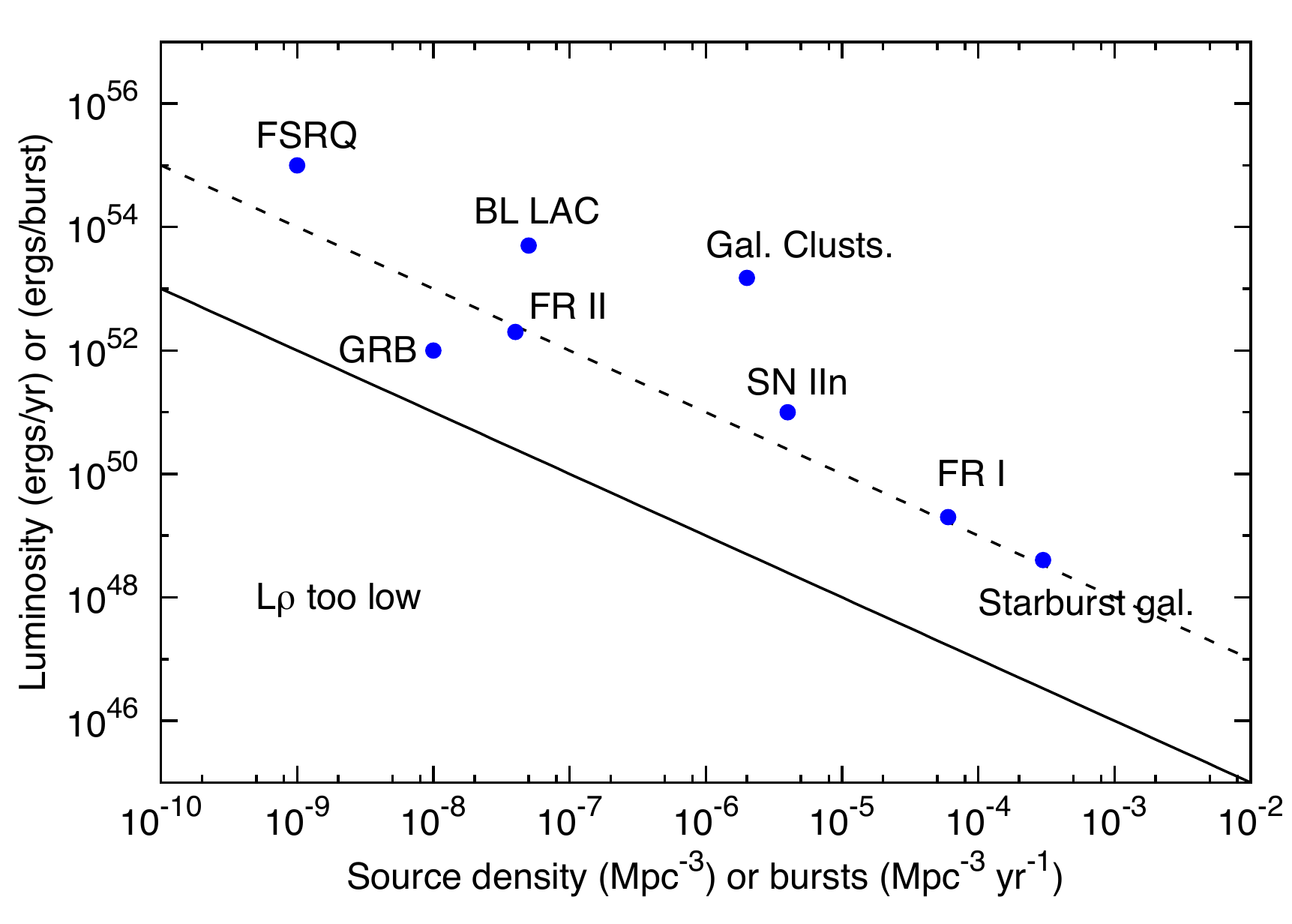}
\caption{Plot of various potential neutrino sources in the plane of luminosity
vs density (after Kowalski~\protect\cite{Kowalski:2014zda}).}
\label{fig:Kowalski}
\end{figure}

The combined constraints on density and luminosity of sources is
displayed in a version of the Kowalski plot~\cite{Kowalski:2014zda}
shown in Fig.~\ref{fig:Kowalski}.  The solid line is the neutrino luminosity
density $Q_\nu\rho_0\sim 10^{43}$~erg/Mpc$^3$ and the broken line
shows the power required for a neutrino efficiency of 1\%.  The 
equivalent analysis for bursting sources plots luminosity per burst vs
density of bursts.  It is interesting that some potential sources
are marginally excluded by the minimum density requirement.
In this situation, multi-messenger searches looking for 
coincidences in various ranges of the electromagnetic
spectrum with neutrinos are of particular interest.

\section{Real Time alerts from IceCube}

In 2016 IceCube set up a real-time alert system~\cite{Aartsen:2016lmt}. Any track-like
event with sufficient energy to have a high probability of being an
astrophysical neutrino generates an alert in the form of a
public GCN circular within a minute of the event.
About 10 alerts have been issued
since the system began.
  
  So far, only
one case~\cite{IceCubeGCN:170922} has led to a coincident observation of
gamma-rays from the same direction, in this case a $\gamma$-ray flare
identified by Fermi~\cite{Fermi:Atel10791} from a blazar TXS 0506+056.
Higher energy $\gamma$-rays were also detected by MAGIC~\cite{MAGIC:Atel10817}.
The significance of these coincident observations is currently being assessed
by all groups with relevant observations.

\section*{Acknowledgment}\vspace{-.3 cm}
I gratefully acknowledge support from the U.S. National Science Foundation 
(PHY 1505990).

\section*{References}\vspace{-.2cm}
\bibliographystyle{utphys}
\bibliography{vietnam}

\end{document}